\documentclass[prl,twocolumn]{revtex4} 
\usepackage{graphicx}
\usepackage{bm}
\usepackage{amsmath, amssymb}

\usepackage{color}
\begin{document}

\title{Finite Temperature Phase Transitions 
in the SU$(N)$ Hubbard model}

\author{Hiromasa Yanatori}
\affiliation{Department of Physics, Tokyo Institute of Technology, 
Tokyo 152-8551, Japan}

\author{Akihisa Koga}
\affiliation{Department of Physics, Tokyo Institute of Technology, 
Tokyo 152-8551, Japan}

\date{\today}%

\begin{abstract}
We investigate the SU($N$) Hubbard model 
for the multi-component fermionic optical lattice system, 
combining dynamical mean-field theory with the continuous-time 
quantum Monte Carlo method. 
We obtain the finite temperature phase diagrams with $N\le 6$ and find that
low temperature properties depends on 
the parity of the components.
The magnetically ordered state competes with the correlated metallic state 
in the system with the even number of components $(N\ge 4)$,
yielding the first-order phase transition.
It is also clarified that, in the odd-component system, 
the ordered state is realized at relatively lower temperatures
and the critical temperature
is constant in the strong coupling limit.
\end{abstract}
\maketitle

Ultracold atomic systems have potential to understand 
some important and fundamental issues in the condensed matter physics~\cite{ultracold1,ultracold2,ultracold3}.
Among them, two-component fermionic systems 
with distinct hyperfine states are known to be appropriate
to describe strongly correlated electron systems.
Owing to the high controllability in particle number, lattice potential, and
interaction strength, remarkable phenomena have been observed such as
superfluid state~\cite{SF1,SF2}, BCS-BEC crossover~\cite{BCSBEC1,BCSBEC2} and 
Mott insulating state~\cite{Mott1,Mott2}.
Recently, the multicomponent fermionic systems are realized such as 
the three components $^6$Li~\cite{6Li}, 
six components $^{173}$Yb~\cite{173Yb} and 
ten components $^{87}$Sr~\cite{87Sr}.
This stimulates further theoretical
investigations on fundamental problems~\cite{3comp1,3comp2}.

One of the interesting systems is the optical lattice system, 
which is realized by loading the ultracold atoms 
in a periodic potential.
This ideal system should be described by the SU($N$) Hubbard model, and 
its ground states have been 
discussed such as the dimerized state 
in the one dimension~\cite{Marston,Kawakami,Assaraf,Buchata},
the staggered flux order in two dimensions~\cite{Marston,Honerkamp},
and some translational symmetry breaking states and 
the superconducting states in the infinite dimensions
~\cite{Okanami,InabaSC,HoshinoWerner,KogaSC,Momoi,Yanatori}.
However, systematic studies for finite temperature properties 
are still lacking~\cite{InabaReview,Yanatori}.
In particular, it is unclear how
the stability of the ordered states depends on 
the parity of the components, 
which should be important to observe the spontaneously translational symmetry
breaking state in the fermionic optical lattice experiments~\cite{1.4T_N}.

Motivated by this, we consider the SU($N$) Hubbard model,
\begin{eqnarray}
\hat{\cal{H}}=-t\sum_{\langle i,j \rangle ,\alpha}
c^{\dagger}_{i\alpha}c_{j\alpha}
+\frac{U}{2}\sum_{i}n_i^2,\label{model}
\end{eqnarray}
where $n_i=\sum_\alpha n_{i\alpha}$ is the total number density of fermions
at the $i$th site, $\langle i,j \rangle$ indicates the nearest neighbor sites
and $c^\dag_{i\alpha}$($c_{i\alpha}$) creates (annihilates) a fermion 
with "color" $\alpha(=1, 2,\cdots, N)$ at site $i$ and 
$n_{i\alpha}=c^\dag_{i\alpha} c_{i\alpha}$.
$t$ is the hopping integral and $U$ 
is the on-site interaction between fermions with distinct components.
Setting chemical potential $\mu=NU/2$,
we discuss the particle-hole symmetric systems.

We examine low temperature properties in the system 
by means of dynamical mean-field theory (DMFT)~\cite{DMFT1,DMFT2,DMFT3}, 
which maps
the lattice model to the problem of a single-impurity connected 
dynamically to an effective medium.
The Green's function is obtained via the self-consistency conditions
imposed of the impurity problem.
We present the fermion band by a semicircular density of state (DOS) 
$\rho(\epsilon)=2\sqrt{1-(\epsilon/D)^2}/(\pi D)$, 
where $D$ is the half-bandwidth.

In the paper, we consider the translational symmetry breaking state
in the bipartite lattice as one of most probable candidates. 
Then, the self-consistent equations~\cite{Chitra} are given by,
${\cal G}_{\gamma\alpha}(i\omega_n)=
i\omega_n+\mu-\left(D/2\right)^2G_{\bar{\gamma}\alpha}(i\omega_n),$
where $G_{\gamma\alpha} (\cal{G}_{\gamma\alpha})$  
is the full (noninteracting) Green function with color $\alpha$ 
for the $\gamma(=A, B)$th sublattice.
To solve the effective impurity problem,
we make use of 
the hybridization-expansion continuous-time quantum Monte Carlo (CTQMC) 
method~\cite{Werner,solver_review},
which should be suitable for systematic investigations on the Hubbard model.

\begin{figure}[htb]
\begin{center}
\includegraphics[width=8cm]{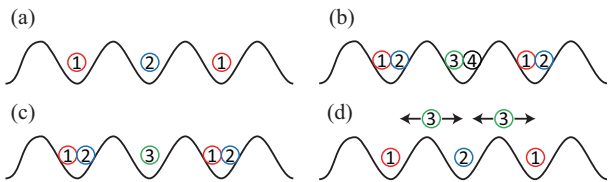}
\caption{(Color online)
(a) AF state for $N=2$. (b) AF state for $N=4$. 
(c) CDW state for $N=3$. (d) CSAF state for $N=3$.
}
\label{fig:pic}
\end{center}
\end{figure}

To discuss how stable the spontaneously translational symmetry
breaking states are at finite temperatures,
we calculate the staggered order parameters 
$m_\alpha=\langle n_{A\alpha}-n_{B\alpha}\rangle/2$,
where $n_{\gamma\alpha}$ is the number operator for color $\alpha$ 
in the $\gamma$th sublattice.
In the system,
the possible ordered states depend on the parity of the components.
When the number of component is even, 
we expect that the repulsive interaction stabilizes 
the "antiferromagnetically" (AF) ordered state, 
where the $N/2$ fermions occupies 
at the $A$ sublattice and the others at the $B$ sublattice.
Schematic pictures for $N=2$ and $4$ 
are shown in Figs.~\ref{fig:pic}(a) and (b).
In the case, the order parameters can be defined as 
$m=m_1=m_2=\cdots=m_{N/2}=-m_{N/2+1}=\cdots=-m_N$.
In the odd component system, some ordered states should be realizable.
For examples, in the three component system, 
the color density wave (CDW) and color-selective antiferromagnetically ordered
(CSAF) states 
are degenerate at zero temperature~\cite{Miyatake},
as shown in Figs.~\ref{fig:pic} (c) and (d).
In the former state, the order parameters have the relation $m_1=m_2\neq m_3$,
while $m_1=-m_2$ and $m_3=0$ in the other.
Therefore, we carefully study the stability of possible ordered states 
in the multicomponent fermionic systems.


Let us consider the system with the even components $(N=2M)$,
which is equivalent to the Hubbard model 
with $M$-fold degenerate bands~\cite{Ono,InabaMott,Blumer}.
We show the order parameters in the systems with $N=2, 4,$ and $6$
at the temperature $T/D=0.02$.
\begin{figure}[htb]
\begin{center}
\includegraphics[width=8cm]{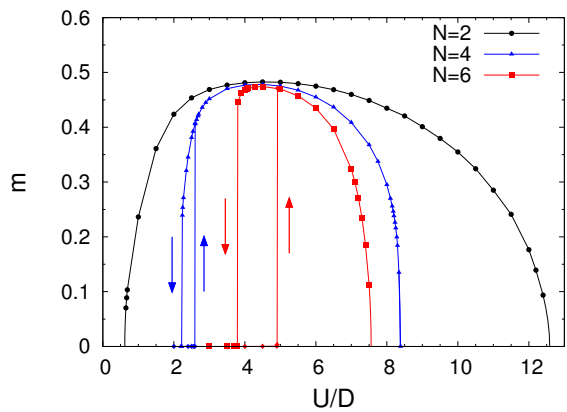}
\caption{(Color online)
Order parameters $m$ as functions of the repulsive interaction $U/D$ 
when $T/D=0.02$. 
Lines are guides to the eye.
The arrows indicate the existence of a hysteresis 
in the order parameters.
}
\label{fig:m_re}
\end{center}
\end{figure}
When $N=2\; (M=1)$, the system is reduced to the single-band Hubbard model.
The normal metallic state is realized in the weak coupling region.
Increasing the repulsive interaction, 
the order parameter is induced at a certain critical interaction $U_{c1}$ and 
the AF state is realized.
Furthermore, increasing the interaction, the order parameter has 
a maximum around $U/D\sim 4$, and decreases.
Finally, it vanishes at another critical interaction $U_{c2}$,
where the second-order phase transition occurs. 
The critical values are deduced as $(U/D)_{c1}\sim 0.62$
and $(U/D)_{c2}\sim 12.6$, 
by examining critical behavior 
$m\sim |U-U_c|^\beta$ with the exponent $\beta=1/2$~\cite{KogaAttractive}.

When $N\ge 4\; (M\ge 2)$, 
the nature of the phase transition in the strong coupling region 
is not changed.
The critical interaction is slightly decreased since
the characteristic energy decreases as $\sim N/(N-1)\cdot D^2/U$.
Namely, these critical values are deduced 
as $(U/D)_c\sim 8.4$ for $N=4$ and $7.6$ for $N=6$. 
However, in the weak coupling region, different behavior appears.
Increasing the interaction, a jump singularity appears in the order parameter
at $(U/D)\sim 2.6\;(N=4)$ and $4.9\; (N=6)$,
as shown in Fig. \ref{fig:m_re}.
On the other hand, decreasing the interaction from the AF state,
the order parameter vanishes at $(U/D)\sim 2.2\;(N=4)$ and $3.8\;(N=6)$.
The hysteresis means the existence of the first-order phase transition
in the even-component system with $N\ge 4$.

\begin{figure}[htb]
\begin{center}
\includegraphics[width=8cm]{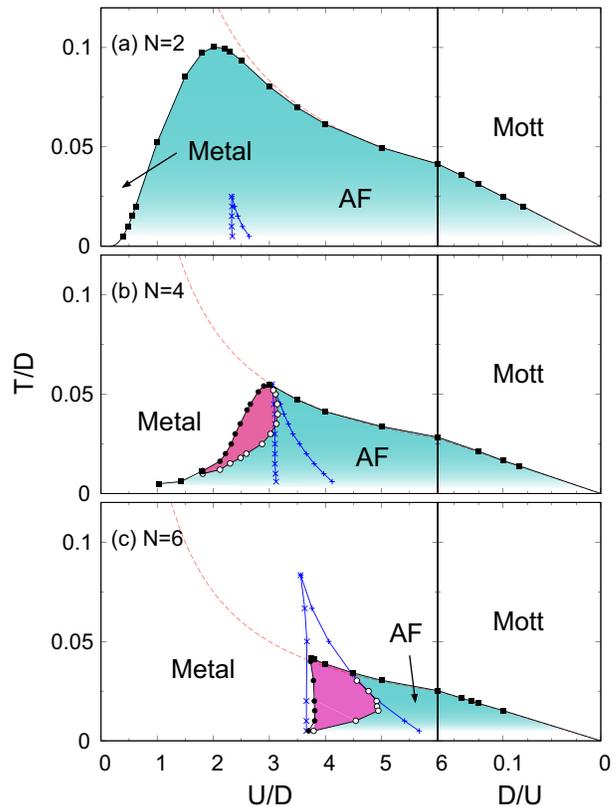}
\caption{(Color online)
Phase diagrams for the multicomponent systems with $N=2, 4,$ and $6$.
Solid squares represent the second-order magnetic phase transition points.
Open (solid) circles represent the transition points, where
the normal (AF) state disappears.
Dashed lines represent the phase boundaries $ND^2/8(N-1)U$
obtained by means of the strong coupling expansion.
Pluses (crosses) represent the transition points 
under the paramagnetic condition,
where metallic (Mott insulating) state disappears.
}
\label{fig:PD_re}
\end{center}
\end{figure}

By performing similar calculations, 
we obtain the finite temperature phase diagrams 
for $N=2, 4$ and $6$, as shown in Fig. \ref{fig:PD_re}.
It is found that in the two-component system, 
the AF state is widely stabilized at finite temperatures.
We also examine low temperature properties under the paramagnetic condition.
Here, we calculate $z=(1-{\rm Im}\Sigma_\alpha(i\omega_0)/\omega_0)^{-1}$
as a renormalization factor at finite temperatures,
where $\Sigma_\alpha$ is the selfenergy for color $\alpha$ 
and $\omega_0=\pi T$.
It is found that the Mott phase boundaries~\cite{Bulla},
where a jump singularity appears in $z$, are much lower than 
the magnetic one.
This is consistent with the fact that
no Mott transitions are realized in the bipartite system
with $N=2$~\cite{Zitzler}.

As increasing $N$, the Mott critical temperature is 
increased~\cite{InabaMott,Blumer}, 
in contrast to the decrease of the magnetic transition temperature.
When $N=4$, the Mott transition temperature is comparable to
the maximum of the magnetic transition, as shown in Fig. \ref{fig:PD_re} (b).
We find that, in the weak and strong couping regions, 
the second-order phase transition occurs.
On the other hand, the first-order phase transition occurs in the intermediate
coupling region $(2\lesssim U/D\lesssim 3)$.
\begin{figure}[htb]
\begin{center}
\includegraphics[width=8cm]{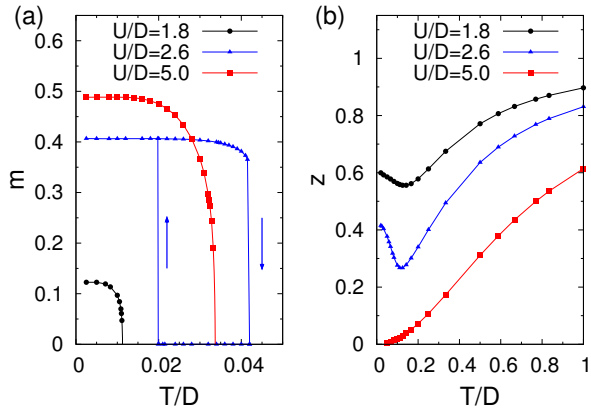}
\caption{(Color online)
Order parameter $m$ (a) and renormalization factor $z$ (b) 
as functions of temperature for the four component systems 
when $U/D=1.8, 2.6$, and $5.0$.
}
\label{fig:T}
\end{center}
\end{figure}
To clarify how particle correlations affect the phase transitions,
we show in Fig.~\ref{fig:T} 
the temperature dependence of the order parameter and renormalization factor.
It is known that the latter is appropriate 
to discuss particle correlations characteristic
of low energy properties.
In the weak coupling region, the renormalization factor is large and 
the normal metallic state is realized 
above the critical temperature $(T/D)\sim 0.011$.
In the AF state ($T<T_c$), the order parameter gradually increases.
In the intermediate coupling region $U/D=2.6$, 
different behavior appears.
Decreasing temperatures, 
the renormalization factor decreases toward zero, implying that 
insulating behavior appears when $T/D>0.2$.
However, it has a minimum around $T/D\sim 0.1$ and 
takes a larger value $z\sim 0.4$ near the first-order transition temperature.
This means the crossover from the insulating state to the metallic state.
It is known that, in the two-band system ($M=1$), 
this correlated metallic state close to the Mott transitions 
is stabilized due to the enhancement of 
spin and orbital fluctuations~\cite{Koga2ED,Koga2QMC}.
At lower temperatures, the jump singularity appears
in the order parameter and the AF state is realized.
In the case, the large gap suddenly appears in the DOS (not shown).
Therefore, in the intermediate coupling region, 
the first-order magnetic transition occurs 
together with the metal-insulator transition.
In the strong coupling region with $U/D=5.0$, 
critical behavior clearly appears around $(T/D)_c\sim 0.034$.
When $T>T_c$, the renormalization factor is small enough to realize the Mott
insulating state.

In the system with $N=6$, 
the Mott critical temperature is higher than the magnetic
transition temperature, as shown in Fig. \ref{fig:PD_re}(c).
Therefore, the Mott transition indeed occurs 
when the interaction strength is changed at the intermediate temperature 
$0.04\lesssim T/D\lesssim 0.08$.
We also find that the metallic state is stable up to a fairly large 
interacting region at finite temperatures.
Then, the phase transition from the metallic (Mott) state to the AF state
is of first (second) order.
These results are essentially the same as those for the $N=4$ system.

In the even-component system with a large $N$, 
the correlated metallic state 
becomes more stable against the magnetic instability.
Roughly speaking, the metallic state is realized in the weak coupling region
until the Mott transition,
which are characterized by three values: the boundaries for 
the coexisting region between metallic and Mott insulating states 
at zero temperature $U_{c1}$ and $U_{c2}$ $\;(U_{c1}<U_{c2})$, and
the critical end point $(U_c, T_c)$.
It is known that $U_{c2}$ is proportional 
to the number of components $N$~\cite{Ono,InabaMott,Blumer}, 
while the others are the square root of $N$~\cite{InabaMott,Blumer}.
On the other hand, the AF state is realized at low temperatures 
$T \lesssim ND^2/8(N-1)U$
in the strong coupling region ($U\gtrsim U_{c2}$).
We conclude that, in the large $N$ case,
the phase diagram is similar to that for the $N=6$ system and
the first-order phase transition between
the correlated metallic and AF states 
occurs in the intermediate coupling region.


Now, we turn to the multicomponent system with $N=2M+1$.
For convenience, we focus on the system with $N=3\;(M=1)$.
It has been clarified that
the CDW and CSAF states, which are schematically 
shown in Figs.~\ref{fig:pic}(c) and (d), are degenerate 
at zero temperature~\cite{Miyatake}.
We have performed the detailed CTQMC calculations
to deduce the critical temperatures for both states,
and little difference between them has been found.
Therefore, in the following, we only show the results for the CSAF state 
to discuss the stability of the ordered states in the system.

\begin{figure}[htb]
\begin{center}
\includegraphics[width=8cm]{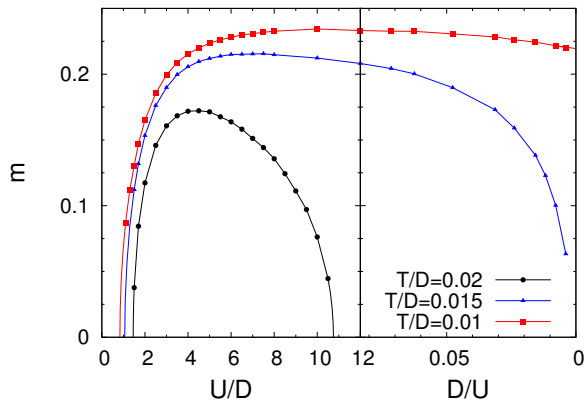}
\caption{(Color online)
Order parameters $m$ as functions of the repulsive interaction $U/D$ 
in the three component system when $T/D=0.01, 0.015$, and $0.02$.
}
\label{fig:m3_re}
\end{center}
\end{figure}

Fig. \ref{fig:m3_re} shows 
the order parameter $m$ in the three-component system.
When $T/D=0.02$, the ordered state is realized 
in the intermediate region $1.5 \lesssim U/D \lesssim 10.8$.
The result is similar to magnetic behavior 
in the two-component system,
where the phase transitions are of second order. 
On the other hand, decreasing temperature, different behavior appears 
in the strong coupling region,
as shown in Fig.~\ref{fig:m3_re}.
When $T/D=0.015$, the increase in the interaction $U$ decreases
the order parameter $m$ slowly. 
We find no phase transition to the paramagnetic state,
at least, until $U/D\le 256$.
At the lower temperature $T/D=0.01$, the order parameter should be finite 
even in the strong coupling limit, 
as shown in Fig.~\ref{fig:m3_re}.

\begin{figure}[htb]
\begin{center}
\includegraphics[width=8cm]{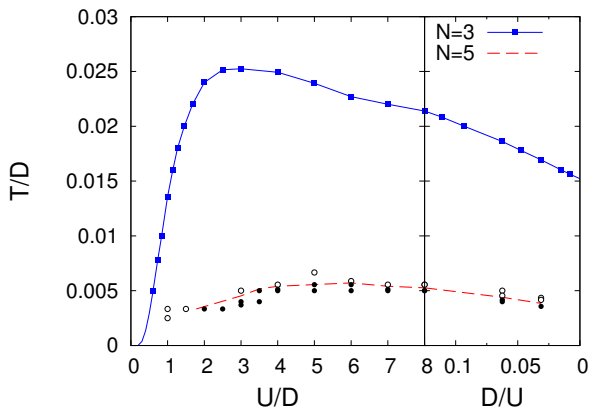}
\caption{(Color online)
Phase diagrams for the odd-component systems with $N=3$ and $5$.
Solid (open) circles indicate the state with (without) the order parameter
for the $N=5$ system.
The phase boundaries are guides to the eyes.
}
\label{fig:PD3_re}
\end{center}
\end{figure}

We obtain the phase diagrams in the systems 
with $N=3$ and $5$, as shown in Fig. \ref{fig:PD3_re}.
In the $N=3$ system, the transition temperature takes 
the maximum $T/D\sim 0.025$ around $U/D=2.5$, 
which is relatively lower than the phase boundaries 
for the even-component system discussed before.
The increase in the interaction strength decreases 
the critical temperature monotonically. 
By extrapolating the phase boundary, we obtain 
a finite critical temperature $(T/D)_c\sim 0.015\;(\sim 1/64)$ 
in the limit $U\rightarrow \infty$.
This remarkable behavior should be explained by the following.
When one considers the CSAF state~[Fig. \ref{fig:pic}(d)], 
a fermion in the doubly occupied sites can hop freely to the nearest neighbor 
singly occupied sites and the metallic state is realized 
even in the strong coupling limit~\cite{Inaba3}.
This is contrast to the even-component system, 
where the corresponding energy cost is proportional to 
the onsite interaction strength $U$.
This should suggest that the effective intersite interaction between 
the localized spins $(S_i^\gamma = \sum_{\alpha,\beta=1,2}c_{i\alpha}^\dag
\sigma_{\alpha\beta}^\gamma c_{i\beta})$ is proportional to
the bandwidth,
resulting in the finite critical temperature $T_c(\propto D)$.
This may recall us the Bose-Einstein condensation 
in a free bosonic system
since the transition temperature does not depend on the interaction strength.
Therefore, it may allow us to discuss the BCS-BEC crossover 
in the two-component Fermi gas,
which is now under consideration.

An ordered state is also realized in the system with $N=5$.
In the case, the transition temperature is much lower than 
that for the $N=3$ system, as shown in Fig. \ref{fig:PD3_re}.
Therefore, we could not determine the phase boundary accurately.
However, the ordered state, shown as solid circles in Fig. \ref{fig:PD3_re},
is realized even in the strong coupling region.
Therefore, we believe that such low temperature properties are
common to the half-filled odd-component system.

We have investigated the phase transitions 
in the multi-component fermion system, 
combining dynamical mean-field theory with 
the hybridization-expansion continuous-time 
quantum Monte Carlo method. 
The finite temperature phase diagrams have been obtained for the Hubbard model 
with $N(\le 6)$ components. 
We have found that 
the magnetically ordered state competes with the correlated metallic state 
in the system with the even number of components.
We have clarified that, in the system with the odd number of components, 
the critical temperature is relatively lower than 
that for the even number cases.
In addition, we have found that the critical temperature
is finite even in the strong coupling limit.

\section*{Acknowledgments}
The authors would like to thank J. Nasu and P. Werner 
for valuable discussions. 
This work was partly supported by the Grant-in-Aid for 
Scientific Research from JSPS, KAKENHI No. 25800193 (A.K.).
The simulations have been performed using some of 
the ALPS libraries~\cite{alps2}.

\end{document}